\documentclass[%
prl%
,twocolumn%
 ,superscriptaddress%
,tightenlines%
,amssymb,amsmath,nobibnotes, aps, prl]{revtex4}
\usepackage{dcolumn}
\usepackage{bm}
\usepackage[dvips]{graphicx}
\begin{document}
\title{NMR Evidence for Spin-Pseudospin Intermixing in Quantum Hall Skyrmions}
\author{Norio Kumada}
\author{Koji Muraki}
\affiliation{NTT\,Basic\,Research\,Laboratories,\,NTT\,Corporation,\,3-1\,Morinosato-Wakamiya,\,Atsugi\,243-0198,\,Japan}

\date{Version: \today}

\begin{abstract}
We investigate quasiparticles in bilayer quantum Hall systems around total filling factor $\nu =1$ by current-pumped and resistively detected NMR.
The measured Knight shift reveals that the spin component in the quasiparticle increases continuously with $\Delta _{\rm SAS}$.
Combined with results for the pseudospin component obtained by activation gap measurements, this demonstrates that both spin and pseudospin are contained in a quasiparticle at intermediate $\Delta _{\rm SAS}$, providing evidence for the existence of the spin-pseudospin intermixed $SU(4)$ skyrmion.
Nuclear spin relaxation measurements show that the collective behavior of the $SU(4)$ skyrmion system qualitatively changes with $\Delta _{\rm SAS}$.
\end{abstract}
\maketitle


In quantum Hall (QH) systems, interactions between electrons often dominate the single-particle physics, leading to rich many-body states with spontaneous symmetry breaking.
This is best illustrated in the QH state at Landau level filling factor $\nu =1$, where an approximate spin $SU(2)$ symmetry is broken by ferromagnetic interactions.
As a consequence of a large Coulomb energy cost of a spin flip, the charged quasiparticle becomes a topological spin texture called a skyrmion, in which spins rotate gradually along the radial direction \cite{Sondhi}.
Experimental evidence for skyrmions has been provided by observing multiple spin flips \cite{Barrett,Schmeller} and low-frequency spin fluctuations \cite{Tycko,Hashimoto} that result from the formation of skyrmions.
While skyrmions are usually associated with the spin degree of freedom, any pseudospin degree of freedom with approximate $SU(2)$ symmetry can host skyrmions.
Indeed, skyrmions originating from the valley degree of freedom have been reported for AlAs quantum wells \cite{Shkolnikov}.
These results motivate one to explore a system with higher $SU(4)$ symmetry, with combined spin and pseudospin degrees of freedom.
Candidates include Si \cite{Arovas} and graphene \cite{Nomura2006,YangSU4,Doretto}, where the valley degree of freedom plays the role of pseudospin, and GaAs/AlGaAs bilayer systems \cite{EzawaSU4_1,EzawaSU4_2,Burkov,Ghosh,Bourassa}, where pseudospin encodes the layer degree of freedom.
Theories predict that in such
systems the quasiparticle is a $SU(4)$ skyrmion (or $CP^3$ skyrmion), in which spin and pseudospin are interwoven.
Furthermore, the interplay between spin and pseudospin in $SU(4)$ skyrmions is expected to lead to a variety of collective modes \cite{Cote2007,Doucot}.

The bilayer system at total filling factor $\nu =1$ can be a good laboratory to search for $SU(4)$ skyrmions.
If the tunneling energy $\Delta _{\rm SAS}$ acting as the pseudospin Zeeman energy is much smaller than the spin Zeeman energy $\Delta _Z$, pseudospin is expected to be the only active degree of freedom \cite{Sarmabook}.
A purely pseudospin origin of the quasiparticle in a system with vanishingly small $\Delta _{\rm SAS}$ has recently been demonstrated by activation gap measurements under a tilted magnetic field \cite{Paula2}.
In the other limit of $\Delta _{\rm SAS}\gg \Delta _{\rm Z}$, the system will become effectively a single layer with spin the only active degree of freedom, where one expects the quasiparticle to evolve into a spin skyrmion \cite{spinskyrmion}.
However, the situation at intermediate $\Delta _{\rm SAS}$ is rather elusive.
Although nuclear spin relaxation measurements have demonstrated that the spin degree of freedom is not completely frozen \cite{KumadaPRL2}, experimental evidence for the $SU(4)$ skyrmion has not been reported yet.

In this Letter, we investigate the quasiparticle around $\nu =1$ in bilayer systems with various $\Delta _{\rm SAS}$.
To study the spin component of the quasiparticle, we exploit current-pumped and resistively detected NMR techniques \cite{KumadaPRL2,KumadaPRL3} and measure the spin depolarization and low-frequency spin fluctuations associated with the quasiparticles.
The extracted number of flipped spins per quasiparticle $N_s$ changes gradually from a large number, $N_s\sim 2.0$ for large $\Delta _{\rm SAS}$, to a value even smaller than unity for small $\Delta _{\rm SAS}$.
Activation gap measurements as a function of $\Delta _{\rm SAS}$ confirm the contribution of the pseudospin degree of freedom in the quasiparticle excitations.
These results, demonstrating that both spin and pseudospin are contained in a quasiparticle at intermediate $\Delta _{\rm SAS}$, provide compelling evidence for the $SU(4)$ skyrmion.
Nuclear spin relaxation measurements reveal the formation of a skyrmion crystal accompanied by gapless collective spin modes at large $\Delta _{\rm SAS}$ and a transition to a qualitatively different state at small $\Delta _{\rm SAS}$.

\begin{table}[b]
\caption{List of the bilayer systems.
Mobility is the value for the total electron density of $1.45\times 10^{11}$\,cm$^{-2}$ in the balanced condition.
$\Delta _{\rm SAS}$ was determined from the Fourier analysis of the low-field Shubnikov-de Haas oscillations.
}
\begin{center}
\renewcommand{\arraystretch}{1.5}
\begin{tabular}{ccc}
\hline\hline
\makebox[15mm]{$\Delta _{\rm SAS}$}  & \makebox[28mm]{Barrier width} & \makebox[31mm]{Mobility (cm$^2$/Vs)}\\
\hline
4\,K & AlAs 1.5\,nm & $1.8\times 10^6$\\
11\,K & Al$_{0.33}$Ga$_{0.67}$As 3.1\,nm\ & $1.6\times 10^6$\\
15\,K & Al$_{0.33}$Ga$_{0.67}$As 2.2\,nm & $1.1\times 10^6$\\
29\,K & Al$_{0.33}$Ga$_{0.67}$As 1.0\,nm & $9.1\times 10^5$\\
\hline\hline
\end{tabular}
\label{sample}
\end{center}
\end{table}

We investigated four samples with different $\Delta _{\rm SAS}$ (Table\,\ref{sample}).
The samples consist of two GaAs quantum wells with the same width of 20\,nm and an Al$_x$Ga$_{1-x}$As tunnel barrier ($x=0.33$ or 1.0) with different thicknesses.
By adjusting the front- and back-gate biases, the filling factors in the front layer $\nu _f$ and in the back layer $\nu _b$ can be controlled independently.
We focus on the balanced-bilayer conditions with equal densities in the two layers ({\it i.e.}, $\nu _f=\nu _b$).
A fixed magnetic field $B=6.0$\,T is applied perpendicular to the two-dimensional plane, corresponding to the constant Zeeman energy $\Delta _Z=1.9$\,K for GaAs ($|g|=0.44$).
At this magnetic field, the ratio between the layer separation and the magnetic length is in the range $2.0\leq d/l_B \leq 2.2$, reflecting the different barrier thicknesses.
For the values of $\Delta _{\rm SAS}$ and $d/l_B$ used here, the system is deep in the QH phase \cite{Murphy}, so that the results reported here are not affected by the phase transition to the compressible state.
Unless otherwise specified, all experiments are carried out at the base temperature $T=50$\,mK.
To apply a radio frequency (rf), a two-turn coil was placed around the sample.
NMR spectra and the relaxation rate were obtained using the current-induced and resistively detected nuclear spin polarization at $\nu _f=2/3$ \cite{KumadaPRL2, KumadaPRL3}.

\begin{figure}[t]
\begin{center}
\includegraphics[width=0.8\linewidth]{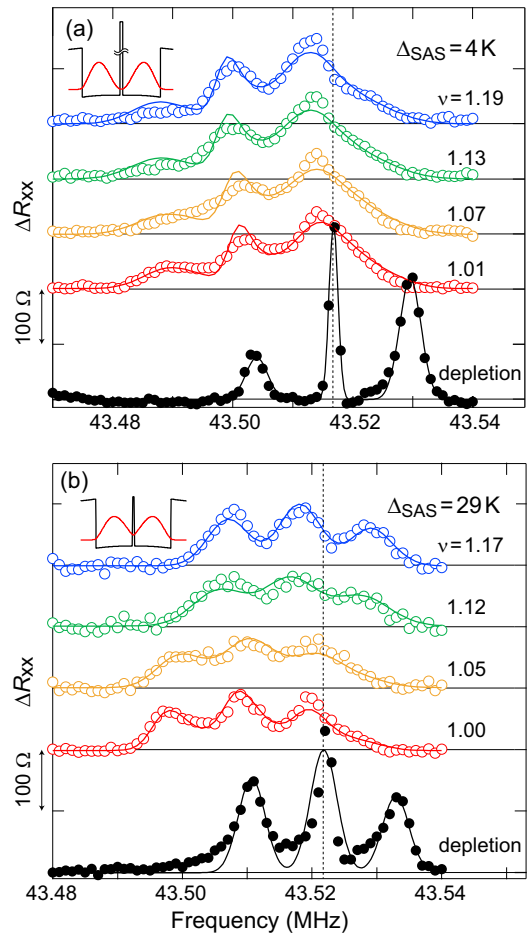}
\caption{(color online).
NMR spectra of $^{75}$As in the balanced condition around $\nu =1$ in the samples with (a) $\Delta _{\rm SAS}=4$\,K and (b) $\Delta _{\rm SAS}=29$\,K.
As a reference, spectra at depletion (solid circles) are shown.
Traces are vertically offset for clarity.
Solid lines are results of fitting.
Dashed vertical lines represent NMR frequency at depletion.
Insets show the calculated potential profile and the distribution of the squared wave function in each sample.
}
\label{spectra}
\end{center}
\end{figure}

\begin{figure}[t]
\begin{center}
\includegraphics[width=0.9\linewidth]{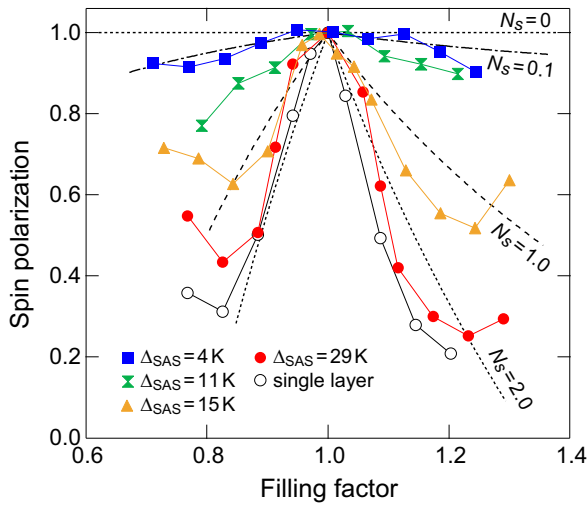}
\caption{(color online).
Spin polarization around $\nu =1$ in the balanced condition (solid symbols).
For comparison, the spin polarization in the single-layer system (open circles) is included.
Dotted and dashed lines represent Eq.\,(\ref{polcal}) for $N_s=0$, 0.1, 1.0 and 2.0.
}
\label{polarization}
\end{center}
\end{figure}


Figure\,\ref{spectra}(a) shows $^{75}$As NMR spectra of the $\Delta _{\rm SAS}=4$\,K sample, taken at various filling factors in the range $1\lesssim \nu \lesssim 1.2$.
The spectra were obtained by measuring the magnitude of the resistance drop $\Delta R_{xx}$ at $\nu _f=2/3$ and $\nu _b=0$ induced by an rf irradiation at each $\nu $ of interest \cite{KumadaPRL3}.
The spectra show three-fold splitting due to the quadrupole interaction induced by sample-dependent unintentional strain.
The spectra are slightly shifted to lower frequency with $\nu $ by the effective magnetic field from electron spins.
The shift, known as the Knight shift $K_s$, is proportional to the product of the spin polarization and electron density.
As a reference, a spectrum taken with all electrons depleted during the rf irradiation is shown.
As shown by the solid lines, these spectra can be well fitted using the procedure described in Ref.\,\cite{KumadaPRL3}, from which $K_s=15.3$\,kHz is obtained for $\nu =1.01$, corresponding to the full polarization.
The $K_s$ increase with $\nu $ can be qualitatively understood as reflecting the increased electron density.

The results for the $\Delta _{\rm SAS}=29$\,K sample are shown in Fig.\,\ref{polarization}(b).
A Knight shift of similar magnitude $K_s=14.6$\,kHz is obtained for $\nu =1.00$, consistent with the full polarization.
On the other hand, strikingly different behavior is observed away from $\nu =1$; the spectra shift to higher frequency with increasing $\nu $.
The corresponding decrease in $K_s$ indicates that the increase in the electron density is overweighed by the decrease in the spin polarization.

The measured $K_s(\nu )$ can be converted into the spin polarization $P(\nu )$ using the relation $P(\nu )/P(\nu =1)=\nu ^{-1}K_s(\nu )/K_s(\nu =1)$, where we assume that $P(\nu =1)=1$.
Figure\,\ref{polarization} compiles the spin polarizations in the range $0.8\lesssim \nu \lesssim 1.2$ for the four samples.
Since increasing (decreasing) the filling factor away from $\nu =1$ results in the creation of quasielectrons (quasiholes), the variation of the spin polarization away from $\nu =1$ reflects the number of spin flips that accompany the introduction of quasiparticles.
For comparison, the result for a single-layer system for $\nu _f\sim 1$ and $\nu _b=0$ is included. The diverse behavior of the four samples shows that the quasiparticles in these systems have largely different characters.
We note that the spatial profiles of the squared wave functions in these systems are similar (insets of Fig.\,\ref{spectra}), which implies similar strength of the Coulomb interactions.
Hence, the different character of the quasiparticles is ascribed to the different magnitude of $\Delta _{\rm SAS}$.

The number of flipped spins $N_s$ in a quasiparticle can be evaluated by fitting the rate of depolarization away from $\nu =1$ using the equation \cite{Khandelwal}
\begin{eqnarray}
P(\nu )=1+2N_s\left( \frac{1}{\nu }-1\right)(\Theta (\nu -1)-\Theta (1-\nu )),
\label{polcal}
\end{eqnarray}
where $\Theta (x)\equiv \{1, x\geq 0;\ 0, x<0\}$.
Comparison with the model shows that the spin component in the $\Delta _{\rm SAS}=29$\,K sample is as large as $N_s\sim 2.0$, which is similar to the value reported (and observed here) for spin skyrmions in the single-layer system \cite{Khandelwal}.
This indicates that, when $\Delta _{\rm SAS}$ dominates $\Delta _Z$, the spin component of the quasiparticle becomes very similar to that of a spin skyrmion in a single-layer system.
As $\Delta _{\rm SAS}$ becomes smaller, the estimated $N_s$ gradually reduces to 1.2, 0.4, and eventually to 0.1 for $\Delta _{\rm SAS}=15$, 11, and 4\,K, respectively.
It is interesting to observe that $N_s$ for $\Delta _{\rm SAS}=11$ and 4\,K are smaller than unity.
If spin is the only active degree of freedom in a quasiparticle, the minimum value of $N_s$ must be unity, corresponding to a single-spin flip \cite{Schmeller}.
The observation of $N_s<1$, in turn, indicates that the quasiparticles must contain the pseudospin degree of freedom.
Notably, even for the smallest $\Delta _{\rm SAS}$, the pseudospin Zeeman energy $\Delta _{\rm SAS}$ ($=4$\,K) is still larger than the spin Zeeman energy $\Delta _Z$ ($=1.9$\,K).
Nevertheless, a pseudospin flip can be energetically cheaper than a spin flip because the interlayer interaction determining the pseudospin exchange energy is smaller than the intralayer interaction determining the spin exchange energy.

\begin{figure}[t]
\begin{center}
\includegraphics[width=0.78\linewidth]{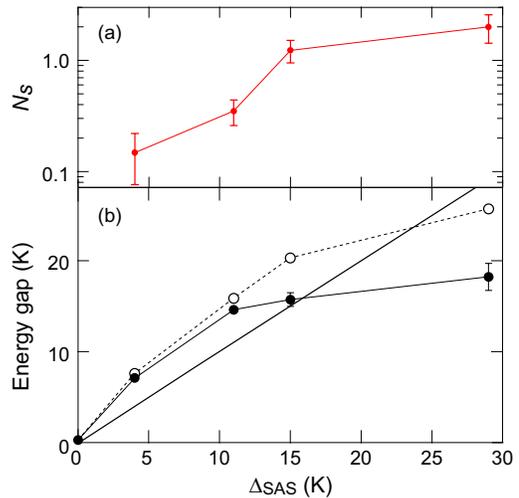}
\caption{(a) $N_s$ obtained by fitting the data in Fig.\,\ref{polarization} using Eq.\,(\ref{polcal}) as a function of $\Delta _{\rm SAS}$.
(b) Activation gap $\Delta $ of the bilayer $\nu =1$ QH state (open circles) as a function of $\Delta _{\rm SAS}$.
The gap $\tilde{\Delta }\equiv \Delta -2N_s\Delta _Z$ (solid circles) with the spin Zeeman term eliminated is also included.
The data point at $\Delta _{\rm SAS}=150$\,$\mu $K is the intrinsic gap of pure pseudospin origin ($N_s=0$) at $d/l_B=2.1$ in Ref.\,\cite{Paula2}.
The solid line represents $\Delta =\Delta _{\rm SAS}$.
}
\label{acti}
\end{center}
\end{figure}

The contribution of the pseudospin degree of freedom in the quasiparticles can be confirmed by measuring the activation gap $\Delta $ for the quasielectron-quasihole excitation as a function of $\Delta _{\rm SAS}$.
In Fig.\,\ref{acti}(b), we plot $\Delta $ (open circles) for the four samples against their respective $\Delta _{\rm SAS}$.
$\Delta $ was determined from the temperature dependence of $R_{xx}$ at $\nu =1$ in the thermally activated regime, where $R_{xx}\propto \exp (-\Delta /2T)$.
The measured $\Delta $ consists of three terms, the Zeeman terms proportional to the numbers of the flipped pseudospins (2$N_p$) and spins (2$N_s$) and the interaction term $\Delta _C$ which also depends on $N_p$ and $N_s$, and can be written in the form
\begin{eqnarray}
\Delta =2N_p\Delta _{\rm SAS}+2N_s\Delta _Z+\Delta _C(N_p,N_s),
\label{delta}
\end{eqnarray}
where the factor 2 accounts for the pair excitation. 
We note here that $N_s$ changes with $\Delta _{\rm SAS}$ as demonstrated in Fig.\,\ref{polarization} and summarized in Fig.\,\ref{acti}(a).
To better elucidate the pseudospin contribution, in Fig.\,\ref{acti}(b) we plot $\tilde{\Delta}\equiv \Delta-2N_s\Delta _Z$ (solid circles) calculated using the values of $N_s$ in Fig.\,\ref{acti}(a).
As $\Delta _{\rm SAS}$ increases, $\tilde{\Delta }$ first increases and then tends to saturate.
The increase at small $\Delta _{\rm SAS}$ demonstrates that the quasielectron-quasihole excitation has a finite pseudospin component, with the slope $\partial\tilde{\Delta} /\partial \Delta _{\rm SAS}$ reflecting 2$N_p$.
On the other hand, the saturation at large $\Delta _{\rm SAS}$ indicates that $N_p$ is approaching zero.
Combined with the results for $N_s$ [Fig.\,\ref{acti}(a)], these results demonstrate the continuous evolution of the quasiparticle from the purely pseudospin excitation (with $N_s=0$) in the small-$\Delta _{\rm SAS}$ limit to the spin skyrmion (with $N_p=0$ and $N_s>1$) in the large-$\Delta _{\rm SAS}$ limit, providing compelling evidence for the $SU(4)$ skyrmion.

\begin{figure}[t]
\begin{center}
\includegraphics[width=1.0\linewidth]{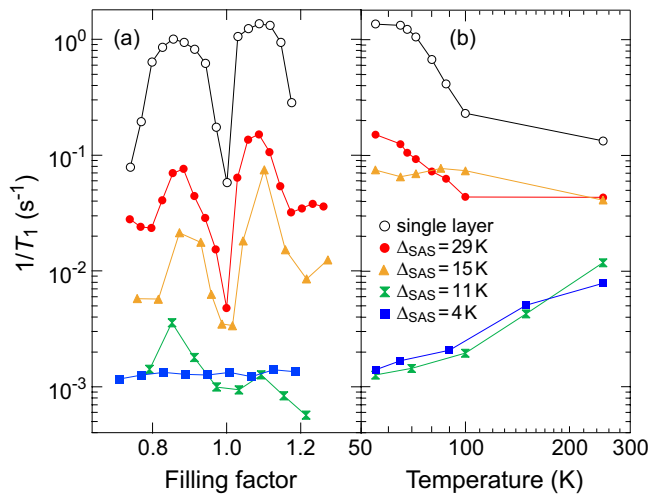}
\caption{(color online).
(a) Nuclear spin relaxation rate $1/T_1$ as a function of the filling factor in the balanced (closed symbols) and single-layer (open circles) conditions.
(b) $1/T_1$ at the peak in (a) near $\nu =1.1$ as a function of the temperature.
}
\label{relaxation}
\end{center}
\end{figure}

The topological nature of the spin component contained in the quasiparticles leads to non-trivial low-frequency spin dynamics and can be investigated through the nuclear spin relaxation rate $1/T_1$, which measures the spectral density of the transverse electron spin fluctuations at the nuclear frequencies ($\sim 50$\,MHz).
Figure\,\ref{relaxation}(a) compares $1/T_1$ in the four samples as a function of $\nu $.
The  $\Delta _{\rm SAS}=29$\,K sample shows behavior very similar to the single-layer system shown as a reference, with $1/T_1$ enhanced on both sides of $\nu =1$ \cite{density}.
Such behavior is well known for spin skyrmions \cite{Tycko,Hashimoto} and is a manifestation of the finite in-plane spin component that results from their non-collinear spin alignment.
As $\Delta _{\rm SAS}$ is reduced, the $1/T_1$ peaks become smaller until they become no longer discernible at $\Delta _{\rm SAS}=4$\,K.
These results are consistent with the change in $N_s$ with $\Delta _{\rm SAS}$.
Namely, as $N_s$ becomes small, the spatial extent of the spin component in the skyrmions shrinks and, accordingly, the in-plane spin magnetization responsible for the enhanced $1/T_1$ reduces.
We note that even at $\nu =1$, $1/T_1$ depends on $\Delta _{\rm SAS}$.
This is presumably due to disorder, which causes local electron densities to deviate from exact $\nu =1$.

Skyrmions interact with each other via the electronic charge and spin they carry and may form a crystal at low temperatures \cite{Brey1995}.
Such a system, called a skyrmion crystal, possesses various collective modes and can support gapless modes \cite{Cote2007}.
The temperature dependence of $1/T_1$ provides information about such collective modes.
In Fig.\,\ref{relaxation}(b), $1/T_1$ at the peak near $\nu =1.1$ is plotted as a function of temperature.
The data reveal three types of behavior: as the temperature is lowered, $1/T_1$ for the large $\Delta _{\rm SAS}$ ($=29$\,K) increases, while that for smaller $\Delta _{\rm SAS}$ ($\leq 11$\,K) decreases. 
For the intermediate $\Delta _{\rm SAS}$ ($=15$\,K), $1/T_1$ is nearly independent of temperature.
The increase in $1/T_1$ with decreasing temperature has been reported for skyrmions in single-layer systems and interpreted as evidence for the formation of a skyrmion crystal \cite{Gervais}.
Such temperature dependence is characteristic of two-dimensional systems, in which the low-frequency fluctuations resulting from the breaking of a continuous symmetry do not freeze out at low temperatures \cite{KumadaScience,Green}.
The result for $\Delta _{\rm SAS}=29$\,K demonstrates the existence of a gapless collective mode.
The opposite temperature dependence for the small $\Delta _{\rm SAS}$ suggests that the spin modes in these systems are gapped.
In this case, $1/T_1$ is governed by thermal fluctuations and thus increases with temperature.
The temperature independence of $1/T_1$ for the intermediate $\Delta _{\rm SAS}$ is then regarded as a signature of a quantum critical point in a second-order phase transition between phases with and without a gapless mode and a spin order \cite{Sachdev}.
These results are qualitatively consistent with a theory \cite{Cote2007} predicting that, as $N_s$ is decreased, the spin order accompanied by the Goldstone modes disappears and a gap opens at a critical $N_s$.

The authors are grateful to L. Tiemann and Z. F. Ezawa for discussions.


\begin{thebibliography}{99}
\bibitem{Sondhi}
S. L. Sondhi, A. Karlhede, S. A. Kivelson, and E. H. Rezayi,
Phys. Rev. B {\bf 47}, 16419 (1993).

\bibitem{Barrett}
S. E. Barrett {\it et al.},
Phys. Rev. Lett. {\bf 74}, 5112 (1995).

\bibitem{Schmeller}
A. Schmeller, J. P. Eisenstein, L. N. Pfeiffer, and K. W. West,
Phys. Rev. Lett. {\bf 75}, 4290 (1995).

\bibitem{Tycko}
R. Tycko {\it et al.},
Science {\bf 268}, 1460 (1995).

\bibitem{Hashimoto}
K. Hashimoto, K. Muraki, T. Saku, and Y. Hirayama,
Phys. Rev. Lett. {\bf 88}, 176601 (2002).

\bibitem{Shkolnikov}
Y. P. Shkolnikov {\it et al.},
Phys. Rev. Lett. {\bf 95}, 066809 (2005).

\bibitem{Arovas}
D. P. Arovas, A. Karlhede and D. Lillieh$\ddot{\rm o}\ddot{\rm o}$k, Phys. Rev. B {\bf 59}, 13147 (1999).

\bibitem{Nomura2006}
K. Nomura and A. H. MacDonald, 
Phys. Rev. Lett. {\bf 96}, 256602 (2006).

\bibitem{YangSU4}
K. Yang, S. Das Sarma, and A. H. MacDonald, 
Phys. Rev. B {\bf 74}, 075423 (2006).

\bibitem{Doretto}
R. L. Doretto and C. M. Smith,
Phys. Rev. B {\bf 76}, 195431 (2007).

\bibitem{EzawaSU4_1}
Z. F. Ezawa, 
Phys. Rev. Lett. {\bf 82}, 3512 (1999).

\bibitem{EzawaSU4_2}
Z. F. Ezawa and G. Tsitsishvili, 
Phys. Rev. B {\bf 70}, 125304 (2004).

\bibitem{Burkov}
A. A. Burkov and A. H. MacDonald,
Phys. Rev. B {\bf 66}, 115320 (2002).

\bibitem{Ghosh}
S. Ghosh and R. Rajaraman, 
Phys. Rev. B {\bf 63}, 035304 (2000).

\bibitem{Bourassa}
J. Bourassa {\it et al.},
Phys. Rev. B {\bf 74}, 195320 (2006).

\bibitem{Cote2007}
R. C$\hat{\rm o}$t$\acute{\rm e}$ {\it et al.},
Phys. Rev. B {\bf 76}, 125320 (2007).

\bibitem{Doucot}
B. Dou\c{c}ot, M. O. Goerbig, P. Lederer, and R. Moessner,
Phys. Rev. B {\bf 78}, 195327 (2008).

\bibitem{Sarmabook}
S. M. Girvin and A. H. MacDonald, 
{\it Perspectives in Quantum Hall Effects} edited by A. Pinczuk and S. Das Sarma (Wiley, New York, 1997).

\bibitem{Paula2}
P. Giudici, K. Muraki, N. Kumada, and T. Fujisawa,
submitted for publication.

\bibitem{spinskyrmion}
We call standard skyrmion ``spin skyrmion'' to distinguish it from ``pseudospin skyrmion''.

\bibitem{KumadaPRL2}
N. Kumada, K. Muraki, K. Hashimoto, and Y. Hirayama,
Phys. Rev. Lett. {\bf 94}, 096802 (2005).

\bibitem{KumadaPRL3}
N. Kumada, K. Muraki and Y. Hirayama,
Phys. Rev. Lett. {\bf 99}, 076805 (2007).

\bibitem{Murphy}
S. Q. Murphy {\it et al.},
Phys. Rev. Lett. {\bf 72}, 728 (1994).

\bibitem{Khandelwal}
P. Khandelwal {\it et al.},
Phys. Rev. Lett. {\bf 86}, 5353 (2001).

\bibitem{density}
Since NMR is performed in the front layer, $1/T_1$ reflects spin fluctuations and the density in the layer.
The quantitative difference between $1/T_1$ in the single-layer system and the bilayer system with $\Delta _{\rm SAS}=29$\,K can be explained by the difference in $\nu _f$: $1/T_1\propto \nu _f^2$.

\bibitem{Brey1995}
L. Brey, H. A. Fertig, R. C$\hat{\rm o}$t$\acute{\rm e}$, and A. H. MacDonald,
Phys. Rev. Lett. {\bf 75}, 2562 (1995).

\bibitem{Gervais}
G. Gervais {\it et al.},
Phys. Rev. Lett. {\bf 94}, 196803 (2005).

\bibitem{KumadaScience}
N. Kumada, K. Muraki, and Y. Hirayama, 
Science {\bf 313}, 329 (2006).

\bibitem{Green}
A. G. Green, 
Phys. Rev. B {\bf 61}, R16299 (2000).

\bibitem{Sachdev}
S. Sachdev, 
Science {\bf 288}, 475 (2000).

\end{thebibliography}
\end{document}